\documentclass[10pt,letterpaper,twocolumn]{article}
\usepackage[english]{babel}
\usepackage{graphicx}

\newcommand{\alt}{\mathbin{\lower 3pt\hbox
   {$\rlap{\raise 5pt\hbox{$\char'074$}}\mathchar"7218$}}}
\newcommand{\agt}{\mathbin{\lower 3pt\hbox
   {$\rlap{\raise 5pt\hbox{$\char'076$}}\mathchar"7218$}}}

\begin{document}

\setcounter{footnote}{0}
\setcounter{equation}{0}
\setcounter{figure}{0}
\setcounter{table}{0}

\title{\large\bf Phase distribution in 1D localization\\
and phase transitions in single-mode waveguides }

\author{\small I. M. Suslov  \\
\small P.L.Kapitza Institute for Physical Problems,  \\
\small 119334 Moscow, Russia  \\
\small E-mail: suslov@kapitza.ras.ru\\
{} \\
\parbox{150mm}{\footnotesize \,Localization of
electrons in 1D disordered systems is usually described
in the random phase approximation, when
distributions of phases $\varphi$ and $\theta$, entering the
transfer matrix, are considered as uniform.
In the general case, the random phase approximation is violated,
and the evolution equations (when the system length $L$ is increased)
contain three independent variables, i.e.
the Landauer resistance
$\rho$ and the combined phases $\psi=\theta-\varphi$
and $\chi=\theta+\varphi$. The phase $\chi$ does not affect
the evolution of $\rho$ and was not considered in previous papers.
The distribution of the phase
$\psi$ is found to exhibit an unusual phase transition at the
point ${\cal E}_0$ when changing the electron energy ${\cal E}$,
which manifests itself
in the appearance of the imaginary part of $\psi$.
The distribution of resistance $P(\rho)$ has no singularity at
the point ${\cal E}_0$, and the transition looks
unobservable in the electron disordered systems.
However, the theory of 1D localization is immediately applicable to
propagation of waves in single-mode optical waveguides.
The optical methods are more efficient and provide possibility
to measure phases $\psi$ and $\chi$.
On the one hand, it makes observable the phase transition in
the distribution $P(\psi)$, which can be considered as a 'trace'
of the mobility edge remaining in 1D systems. On the other hand,
observability of the phase $\chi$ makes  actual
derivation of its evolution equation, which is
presented
below. Relaxation of the distribution
$P(\chi)$ to the limiting distribution $P_\infty(\chi)$ at $L\to\infty$
is described by two exponents, whose exponentials have jumps of
the second derivative, when the energy ${\cal E}$ is changed.  }
}

\date{}
\maketitle


\setcounter{footnote}{0}
\setcounter{equation}{0}
\setcounter{figure}{0}
\setcounter{table}{0}

\begin{center}
{\bf 1. Introduction}
\end{center}

Localization of electrons in 1D disordered systems
can be conveniently described using the transfer matrix
$T$, relating the amplitudes of plane waves on the left
($Ae^{ikx}+Be^{-ikx}$) and on the right ($Ce^{ikx}+De^{-ikx}$) of
a scatterer,
$$
\left ( \begin{array}{cc} A \\ B \end{array} \right)\,
=  T \left ( \begin{array}{cc} C \\ D \end{array}\right)
\,.
\eqno(1)
$$
In the presence of the time-reversal invariance, the matrix $T$
can be parametrized in the form \cite{1}
$$
 T= \left ( \begin{array}{cc} \!\!\! 1/t\! &\! - r/t \!\!\\
\!\!- r^*/t^* \!&\! 1/t^* \!\!\!\end{array} \right)\,
= \left ( \begin{array}{cc}
\!\!\sqrt{\rho\!+\!1}\, e^{i\varphi}\!\! &
\!\!\sqrt{\rho} \,e^{i\theta}\!\!
\\ \!\!\sqrt{\rho}\, e^{-i\theta}\!\!
&\!\! \sqrt{\rho\!+\!1}\,
e^{-i\varphi}\!\! \end{array} \right)\,,
\eqno(2)
$$
where $t$ and $r$ are the amplitudes of transmission
and reflection, while $\rho=|r/t|^2$ is the dimensionless
Landauer resistance \cite{2}.
 For the successive
arrangement of scatterers their transfer matrices are
multiplied. For a weak scatterer its transfer matrix $T$ is
close to the unit one, allowing one to derive the differential
evolution equations for its parameters.

Usually, such equations are derived in the random phase
approximation, when distributions of $\varphi$ and
$\theta$ are considered as uniform  \cite{3}--\cite{8}.
Such approximation is working
sufficiently good for weak disorder
in the deep
of the allowed band, as it is usually accepted
in theoretical papers (see references in \cite{9,10,11}).
The fluctuation states in the forbidden band are
considered infrequently \cite{12,13,14} and only on the level of
wave functions. A systematic analysis shows that the random phase
approximation is strongly violated near the
initial band edge and in the forbidden band of an
ideal crystal \cite{15}. In the general case, the evolution
equations are written in terms of the Landauer resistance
$\rho$ and the combined phases (Sec.2)
$$
\psi=\theta-\varphi\,,\qquad
\chi=\theta+\varphi\,.
\eqno(3)
$$
The phase $\chi$ does not affect the evolution of $\rho$
and is not interesting for the condensed matter physics;
so it was not discussed in the previous papers \cite{15,16,17}.
Optical measurements (see below) allow to study the distribution
of the phase $\chi$, and its theoretical investigation becomes
actual.

The complete evolution equation for the distribution
$P(\rho,\psi,\chi)$ is derived in Appendix. In fact, it has no
practical value, and only its general structure is essential,
which allows separation of variables (Sec.2).
Factorization  $P(\rho,\psi,\chi)=P(\rho,\psi)P(\chi)$
is valid for an arbitrary system length $L$, and allows to confine
oneself to equations for $P(\rho,\psi)$ and $P(\chi)$.
Additional factorization $P(\rho,\psi)=P(\rho)P(\psi)$
arises for large $L$ and leads to the closed equation for
$P(\rho)$ and the equation for the stationary distribution
$P(\psi)$.

The stationary distribution of the phase $\psi$ was studied
in the papers \cite{16,17}; in the deep of the disordered system
it undergoes the peculiar phase transition in the point ${\cal E}_0$,
when the electron energy ${\cal E}$ is changed \cite{17}, consisting
in appearance of the imaginary part of $\psi$ (Sec.3).
Meanwhile, the distribution of resistance $P(\rho)$ has no
singularity at the point ${\cal E}_0$, and the transition  looks
unobservable in the framework of
condensed matter physics.

The  evolution equation for $P(\chi)$ is derived in Sec.4: it has
a form of the usual diffusion equation, where the diffusion constant
and drift velocity are exponential functions of $L$. The
corresponded exponentials have singularities as functions of
${\cal E}$, consisting in jumps of the second derivative (Sec.5).
Such phase transitions are also unobservable for the electron
disordered systems.

However, the approach developed previously \cite{15,16,17} is
immediately applicable to the scattering
of waves propagating in
single-mode optical waveguides (Sec.6.1).
Existent optical methods (heterodyne approach, near-field
microscopy, etc.) are rather efficient, and
allow to measure distributions of all parameters
$\rho$, $\psi$, $\chi$ inside the
waveguide\,\footnote{\,In this context,
the parameter $\rho$ does not
have a meaning of
Landauer resistance, but determines the amplitudes of
transmitted and reflected waves (Sec.6.2).} (Sec.6.3).
It extends the observable aspects of the 1D localization
theory, and provides possibilities for its deep experimental
verification. In particular, the phase transitions in
distributions $P(\psi)$ and $P(\chi)$ become observable
(Secs.6.2, 6.3). The possible schemes of measurement are
described in Sec.6.4.

The brief communication on the obtained results was
made previously by the author and S.\,I.\,Bozhevolnyi \cite{300}.

\begin{center}
{\bf 2. General structure of evolution equations }
\end{center}

\begin{figure*}
\centerline{\includegraphics[width=5.0 in]{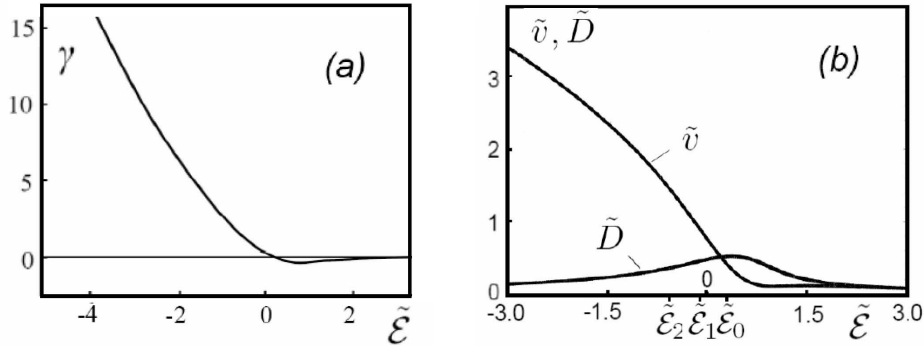}}
\caption{\small Dependence of parameters $\gamma$,
$\tilde v=v/W^{2/3}$ and $\tilde D=D/W^{2/3}$ on the reduced
energy $\tilde{\cal E}={\cal E}/W^{4/3}$, obtained from the
analysis of moments for the transfer matrix elements
\cite{15}. These moments are regular functions of energy,
which leads to regularity of the presented dependencies.
Smallness of $\gamma$ and the equality $v=D$, valid in the random
phase approximation, are realized only in the deep of the allowed
band. The points $\tilde{\cal E}_0$,   $\tilde{\cal E}_1$,
$\tilde{\cal E}_2$ correspond to phase transitions, discussed in
Secs.3,\,5. }
\label{fig1} \end{figure*}

The most general evolution equation describes the change
of the mutual distribution $P(\rho,\psi,\chi)$ under
increasing of the system length $L$ and has the following
structure (see Appendix)
$$
\frac{\partial P}{\partial L}=
\left\{\vphantom{L^2} \hat L_{\rho,\psi} P \right\}'_\rho +
\left\{ \vphantom{L^2} \hat M_{\rho,\psi} P \right\}'_\psi \,
+\left\{ \vphantom{L^2} \hat K_{\rho,\psi,\chi} P \right\}'_\chi
\,,
\eqno(4)
$$
where $\hat K$, $\hat L$, $\hat M$ are operators, depending on
indicated variables.
The right-hand side is the sum of full derivatives, which
ensures the conservation of probability. As was discussed in
\cite{17,18}, conditions for separation of variables in the
diffusion-type equations are essentially weaker, than for an
eigenvalue problem.  Independence of $\chi$ for operators $\hat
L$ and $\hat M$ provides factorization
$P(\rho,\psi,\chi)=P(\rho,\psi)P(\chi)$, where $P(\rho,\psi)$ and
$P(\chi)$ are determined by equations
$$
\frac{\partial P(\rho,\psi)}{\partial L}=
\left\{\vphantom{L^2} \hat L_{\rho,\psi} P(\rho,\psi) \right\}'_\rho +
\left\{ \vphantom{L^2} \hat M_{\rho,\psi} P(\rho,\psi) \right\}'_\psi
\eqno(5)
$$
and
$$
\frac{\partial P(\chi)}{\partial L}=
\left\{\vphantom{L^2} \hat {\cal K}_{\chi} P(\chi)
\right\}'_\chi \,, \quad
\hat {\cal K}_{\chi} = \int \hat K_{\rho,\psi,\chi}
P(\rho,\psi)\,d\rho\,d\psi \,.
\eqno(6)
$$
The specific form of Eq.5 is given in \cite{16,17}, while
Eq.6 is derived in Sec.4. In the large
$L$ limit, when the typical values of $\rho$ are large, the
operator $\hat M_{\rho,\psi}$ becomes independent of $\rho$;
then the solution of Eq.5 is factorized,
$P(\rho,\psi)=P(\rho) P(\psi)$, where $P(\rho)$ and $P(\psi)$
are determined by equations
$$
\frac{\partial P(\psi)}{\partial L}=
\left\{\vphantom{L^2} \hat M_{\psi} P(\psi)
\right\}'_\psi \,,
\eqno(7)
$$
$$
\frac{\partial P(\rho)}{\partial L}=
\left\{\vphantom{L^2} \hat {\cal L}_{\rho} P(\rho)
\right\}'_\rho \,, \quad
\hat {\cal L}_{\rho} = \int \hat L_{\rho,\psi}
P(\psi)\,d\psi \,.
\eqno(8)
$$
Equation (7) provides the existence of the stationary
distribution of the phase $\psi$.
The equation (8) for $P(\rho)$ has a form \cite{15}
$$
\frac{\partial P(\rho)}{\partial L} =
D\,\frac{\partial}{\partial \rho}
\left[\,-\gamma(1\!+\!2\rho) P(\rho) +
\rho(1\!+\!\rho)\,\frac{\partial P(\rho)}{\partial \rho}
\,\right]   \,,
\eqno(9)
$$
and gives at large $L$ the limiting log-normal distribution
$$
P(\rho)=\frac{1}{\rho \sqrt{4\pi D L}}
\exp\left\{-\frac{[\ln \rho-vL]^2}{4DL}\right\}\,
\eqno(10)
$$
with $v=(2\gamma\!+\! 1)D$. The typical value of $\rho$
increases exponentially with  $L$, which
is an observable manifestation of 1D localization.
In the random phase approximation, the parameter $\gamma$
turns to zero, and equations (9), (10) coincide with
the previously obtained results
\cite{3}--\cite{8}. Dependencies of $\gamma$,
$D$, $v$ on the reduced energy $\tilde{\cal E}={\cal E}/W^{4/3}$,
obtained from the analysis of moments for distribution of
the transfer matrix elements \cite{15},
are shown in Fig.1, where ${\cal E}$ is the energy counted from
the initial band edge, and $W$ is an amplitude of the random
potential; all energies are measured in units of the hopping
integral of the 1D Anderson model, which is of the order of
the initial band width.
Strong violation of the random phase
approximation is thereby evident (Fig.1).

It should be clear that the specific form of Eq.4 is of no
importance, and only its general structure is essential.
For arbitrary $L$, Eq.4 is split to two equations (5) and
(6), while for large $L$ it reduces to three equations (6), (7), (8).
It is also clear, that the choice of independent variables
$\rho$, $\psi$, $\chi$ has the objective character.

\begin{center}
{\bf 3. Phase transition in the distribution $P(\psi)$ }
\end{center}

The meaning of the phase transition in the $\psi$ distribution
consists in the fact that difference between the allowed
and forbidden bands survives (in a certain sense) in the
presence of the random potential, though a singularity in
the density of states is smoothed out. It resembles the famous
argumentation by Mott \cite{19}, that
the role of the allowed band edge comes to the mobility
edge. Although the mobility edge is absent in the 1D case,
a 'trace' of it still remains. The point is that
the probe scatterer in the allowed band (${\cal E}>0$) is
described by the transfer matrix (2), while in the forbidden band
(${\cal E}<0$) it is described by the pseudo-transfer matrix
${\cal T}$ \cite{15}, relating coefficients of the increasing and
decreasing exponents on the left ($Ae^{\kappa x}+Be^{-\kappa x}$)
and on the right ($Ce^{\kappa x}+De^{-\kappa x}$) of the
scatterer.  In the simplest case, the matrix ${\cal T}$ is real
and corresponds to pure imaginary values of phases $\theta$ and
$\varphi$. Let us compare situations for ${\cal E}>0$ and
${\cal E}<0$: for a sufficient separation
in energy, the difference  between two types of
matrices can be made arbitrary large, and it cannot be
overcome by addition
of weak disorder.
As a result, the border-line between the true and
pseudo transfer matrices can only be shifted, but not
eliminated\footnote{\,One can object, that existence of a random
potential violates spatial homogenity, and a shift
of the border-line becomes dependent on the position of the
probe scatterer, leading to smearing of the phase
transition. Physically, it is so indeed, and this is a
reason for regularity of the Landauer resistance $\rho$.
However, the indicated band edge fluctuations
correspond to the spatial fluctuations of the phase $\psi$.
The crucial point is that the distribution $P(\psi)$ is
stationary and obeys spatial homogenity in the deep
of the system: it is determined by a set of parameters, which
are independent of the coordinate.
Consequently, for the whole $\psi$ distribution  the border-line
between true and pseudo transfer matrices lies at a strictly
defined energy. The stationary distribution $P(\psi)$ appears to
be the same both for a change of the coordinate for a specific
configuration of the potential, and for a change of its
realization: in fact, it is usual ergodicity, since the
coordinate $x$ (Sec.6) plays a role of a time. }.
In practice, it is manifested
via the appearance of the imaginary part of the phase $\psi$
for energies ${\cal E}<{\cal E}_0$ \cite{17}.

The formal statements of the paper \cite{17} reduce to
the following. First of all, one should differ the
'external' and 'internal' phase distributions (Fig.2).
The internal phase distribution is realized in the deep of a
sufficiently long disordered system, and is independent of
boundary conditions. Considering the system from
the side of ideal leads, one observes the 'external' phase
distribution, which is determined by the boundary conditions;
namely these phases appear in the transfer matrix. Influence of
interfaces extends till the length scale of the order of the
localization length $\xi$: it determines the transient
region, where the internal phase distribution continually
transforms to the external one.  In the large $L$ limit, the
distribution $P(\rho)$ is determined by the internal phase
distribution, which provides its independence from the
boundary conditions. However, the evolution equations contain
namely the external phase distribution, and one wonders
why it does not affect the
limiting distribution $P(\rho)$. The second question, related
with the first one, is as follows: how can we find the internal
phase distribution, if it does not appear in the evolution
equations?

The above questions are resolved in the following
manner. The phase $\psi$ appears to be a 'bad' variable,
while the 'correct' variable is
$$
w=-{\rm cot}\,\psi/2 \,.
\eqno(11)
$$
The form of the stationary distribution $P(w)$ is
determined by the internal properties of the system
and does not depend on the boundary conditions.
If the boundary conditions are changed, it leads to three
effects: the scale transformation $w\to sw$ and
two translations
$w \to w+w_0$ and $\psi\to\psi+\psi_0$. The corresponding
changes of the distribution $P(\psi)$ are easily
predictable  \cite{17} and can be observed in the external
phase distribution. The evolution equations are invariant
in respect to translation $\psi\to\psi+\psi_0$, and the internal
phase distribution can be discussed at some fixed choice of
the origin. Invariance of the limiting distribution $P(\rho)$
under transformations $w\to sw$ and $w\to w+w_0$
is realized in the dynamical manner. Analogously to
aperiodic oscillations of $P(\rho)$ \cite{122,123},
in the region $L\alt \xi$
the scale factor $s$ and the translational shift $w_0$
undergo aperiodic oscillations as functions of $L$,
attenuating at large $L$. As a result, $s$ and $w_0$ tend
to the certain 'correct' values, which provide the
correct values of $D$ and $v$ in the limiting distribution
 (10). The indicated 'correct' values\,\footnote{\,A
 meaning of these values of $s$ and $w_0$ consists in the fact
 that the distribution  $P(\psi)$ becomes stationary
 only for certain 'correct' boundary conditions, which
 are formed automatically at a distance of order $\xi$ from the
 ends of the system. If such 'correct' boundary conditions
 (specified by $s$ and  $w_0$) are chosen at the ends of
 the system, then the transient region of the order of $\xi$
 dissappears, and the stationary distribution is formed at very
 small scales: as a result, the difference between the
 'external' and 'internal' phase distributions  (Fig.2)
 practically dissappears. It gives the way to establish
 the 'internal' phase distribution, which is not contained
 in the evolution equations, through the 'external' phase
 distribution, entering these equations.  } correspond to the
  internal phase distribution, and the latter can be found after
 return to the variable $\psi$. Meanwhile, it appears that the
translational shift $w_0$ becomes complex-valued for
${\cal E}<{\cal E}_0$, indicating  the appearance
of the imaginary part of $\psi$. This qualitative change
indicates the existence of the unusual phase transition.

The point ${\cal E}_0$ is not singular for the Landauer
resistance  $\rho$, and the whole distribution $P(\rho)$
varies in its vicinity in a smooth manner  (Fig.1,b). As a
result, the described phase transition looks
unobservable in the framework of the condensed matter physics.
Fortunately, it has the observable manifestations in
optics  in the form of the square
root singularities in the frequency dependencies (Secs.6.2,6.3).

\begin{figure}
\centerline{\includegraphics[width=3.1 in]{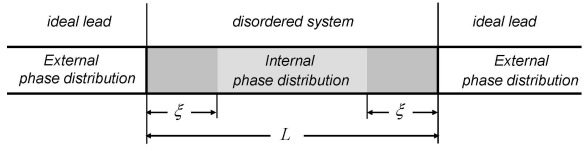}}
\caption{\small External and internal phase distributions.
}
\label{fig2}
\end{figure}

\begin{center}
{\bf 4. Evolution equation for $P(\chi)$ }
\end{center}

According to  \cite{17}, the change of the transfer matrix
$T^{(n)}$ under increasing the number of scatterers $n$ is
determined by the recurrence relation
$$
T^{(n+1)}=T^{(n)}T_\delta T_{\epsilon_n} \,,
\eqno(12)
$$
where matrices  $T^{(n)}$ and $T_{\epsilon_n}$  are statistically
independent, and $T_\delta$ is
constant. These matrices can be accepted in the form
$$
 T_{\epsilon_n}=
\left ( \begin{array}{cc} 1\!-\!i\epsilon_n
& \epsilon_n {\rm e}^{i\gamma} \\
\epsilon_n {\rm e}^{-i\gamma} &
1\!+\!i\epsilon_n \end{array} \right)\,,\qquad
\eqno(13)
$$
$$
 T_{\delta}= \left ( \begin{array}{cc}
{\cal A} & {\cal B} \\ {\cal B}^* & {\cal A}^* \end{array} \right)\,=
\left ( \begin{array}{cc}
\sqrt{1\!+\!\Delta^2}\, {\rm e}^{i\alpha} \!\!  &
\Delta {\rm e}^{i\beta} \!\!\!
\\ \Delta {\rm e}^{-i\beta} \! & \sqrt{1\!+\!\Delta^2}\, {\rm
e}^{-i\alpha}\!\! \end{array} \right).
\eqno(14)
$$
where $\epsilon_n$ is proportional to the
amplitude of the $n$th scatterers, and
$\langle\epsilon_n\rangle=0$,
$\left\langle\epsilon^2_n\right\rangle\equiv \epsilon^2$,
while $T_{\delta}$ is determined by the parameter $\delta$,
proportional to the distance between
scatterers\,\footnote{\,The constancy of $T_{\delta}$
takes place, if the distances between scatterers are
equal. For example, in the 1D Anderson model a scatterer is
present in each site of the lattice: in this case the number of
scatterers $n$ coincides with the system length $L$ in units
of the lattice constant.  }, so that  $\Delta\sim
\alpha\sim \delta$ \cite{17}. Below we consider the limit
$$
\delta\to 0\,, \quad \epsilon
\to 0\,,\quad \delta/\epsilon^2=const
\eqno(15)
$$
and retain the terms of the first order in  $\delta$ and the
second order in  $\epsilon$.

Accepting parametrization (2) for $T^{(n)}$ and denoting
parameters of $T^{(n+1)}$ as $\tilde \rho$, $\tilde \varphi$,
$\tilde \theta$, we have
$$
\sqrt{1\!+\!\tilde\rho}\,{\rm e}^{i\tilde\varphi}
= \sqrt{1\!+\!\rho}\, {\rm e}^{i\varphi}
({\cal A}+\epsilon {\cal C}) +\sqrt{\rho}\, {\rm e}^{i\theta}
({\cal B}^*\!+\epsilon {\cal D}^*) \,,
\eqno(16)
$$
$$
\sqrt{\tilde\rho}\,{\rm e}^{i\tilde\theta}
= \sqrt{1\!+\!\rho}\, {\rm e}^{i\varphi}
({\cal B}\!+\!\epsilon {\cal D}) +\sqrt{\rho}\, {\rm e}^{i\theta}
({\cal A}^*\!+\epsilon {\cal C}^*) \,,
$$
where the following notations are accepted
$$
{\cal C}=\!{\cal B}\,{\rm e}^{-i\gamma}\!-\!i{\cal A}\,
\qquad
{\cal D}=\!{\cal A}\,{\rm e}^{i\gamma}\!+\!i{\cal B}\,.
\eqno(17)
$$
Squaring in modulus one of equations (16) and omitting index
of $\epsilon_n$, we have
$$
\tilde\rho= \rho+{\cal K} \sqrt{\rho(1\!+\!\rho)}
+\epsilon^2 (1\!+\!2\rho)\,, \
\eqno(18)
$$
where
$$
{\cal K}= 2\Delta\cos{(\psi\!\!-\!\!\beta)} +2\epsilon
\cos{(\psi\!\!-\!\!\gamma)} -2\epsilon^2
\sin{(\psi\!\!-\!\!\gamma)}\,.
\eqno(19)
$$
Taking the product of the second equation (16) with the
complex-conjugated first equation, and excluding
$\tilde\rho$ with the help of Eq.18, one obtains the
relation between $\tilde\psi$ and $\psi$ \cite{17}
$$
\tilde\psi=\psi+2\,(\epsilon\!-\!\alpha)+
(R^2/2\!-\!1)\,\epsilon^2\sin{2(\psi\!-\!\gamma)}-
$$
$$
-R\,\left[\Delta\sin{(\psi\!-\!\beta)}
+\epsilon\sin{(\psi\!-\!\gamma)}
+\epsilon^2\cos{(\psi\!-\!\gamma)} \right]\,,
\eqno(20)
$$
where
$$
R=\frac{1\!+\!2\rho}{\sqrt{\rho(1\!+\!\rho)}} \,.
\eqno(21)
$$
Equations (18) and (20) allow to derive the evolution
equation (5) for $P(\rho,\psi)$ \cite{17}. Now let take
the product of two equations (16)
$$
\sqrt{\tilde\rho(1\!+\!\tilde\rho) }\,
{\rm e}^{i\tilde\chi-i\chi}=
\sqrt{\rho(1\!+\!\rho)}\,(1+2\epsilon^2)+
$$
$$
+\Delta\left[ {\rm e}^{i(\beta-\psi)}+2\rho\cos{(\beta\!-\!\psi)} \right]
$$
$$
+\epsilon\left[ {\rm e}^{i(\gamma-\psi)}+2\rho\cos{(\gamma\!-\!\psi)}
\right]-
$$
$$
-\epsilon^2\left[ i{\rm
e}^{i(\gamma-\psi)}-2\rho\sin{(\gamma\!-\!\psi)} \right]
\eqno(22)
$$
and excluding  $\tilde \rho$, find the relation between
$\tilde\chi$ and $\chi$
$$
\tilde\chi=\chi-f(\rho,\psi)\,,
$$
$$
f(\rho,\psi)=\frac{\! \Delta\sin{(\psi\!-\!\beta)}\!+\!
\epsilon\sin{(\psi\!-\!\gamma)}\! +\!\epsilon^2\cos{(\psi\!-\!\gamma)}\! }
{\sqrt{\rho(1\!+\!\rho)}} -
$$
$$
-\frac{\epsilon^2(1\!+\!2\rho)\sin{2(\psi\!-\!\gamma)}}{2\rho(1\!+\!\rho)}\,.
\eqno(23)
$$
The evolution equation for $P(\chi)$ is composing according to
the rule
$$
P_{n+1}(\tilde\chi)=\int
\delta\left(\vphantom{L^2}\tilde\chi-\chi+f(\rho,\psi)\right)
P_n(\chi) \cdot
$$
$$ \cdot
P_n(\rho,\psi) P_n(\epsilon) \, d\chi\, d\rho\, d\psi\,
d\epsilon  \,,
\eqno(24)
$$
and accepts the following form after trivial integration over
$\chi$
$$
P_{n+1}(\chi)=\left\langle \vphantom{L^2}
    P_n\left(  \vphantom{L^2} \chi+f(\rho,\psi) \right)
	 \right\rangle \,,
\eqno(25)
$$
with averaging over $\rho$, $\psi$, $\epsilon$.
Expanding the right-hand side over the small
increment $f(\rho,\psi)$, one has
$$
P_{n+1}(\chi)-P_{n}(\chi)=
\left\langle \vphantom{L^2} f(\rho,\psi)\right\rangle
\frac{d P_n}{d\chi}
+\frac{1}{2} \left\langle f(\rho,\psi)^2\right\rangle
\frac{d^2 P_n}{d\chi^2}   \,,
\eqno(26)
$$
which leads to the final equation
$$
\frac{\partial P}{\partial L} =-v^* P'_\chi +
D^*P''_{\chi \chi}
\eqno(27)
$$
having a form of the usual diffusion equation with
variables coefficients
$$
v^*\!=\!\left\langle \vphantom{L^2}
\frac{\! -\Delta\sin{(\psi\!-\!\beta)}\!
 +\!\epsilon^2\cos{(\psi\!-\!\gamma)}
 \left[ R\sin{(\psi\!-\!\gamma)} -1  \right]\! }
{\sqrt{\rho(1\!+\!\rho)}}  \right\rangle  \,,
$$
$$
D^*=\left\langle \vphantom{L^2}
\frac{\! \epsilon^2\sin^2{(\psi\!-\!\gamma)}\! }
{2\rho(1\!+\!\rho)}
   \right\rangle   \,,
\eqno(28)
$$
which are determined by averages over the
distribution $P(\rho,\psi)$.

\begin{center}
{\bf 5. Phase transitions in distribution $P(\chi)$ }
\end{center}

The typical values of $\rho$ are large for large $L$,
and the main order in $1/\rho$ is sufficient in Eq.28.
In addition, the distribution $P(\rho,\psi)$ is
factorized and one has independent averaging over $\rho$
and $\psi$:
$$
v^*\!\!=\!\left\langle \vphantom{L^2}
\!\! -\!\Delta\sin{(\psi\!-\!\beta)}\!
 -\!\epsilon^2 \!\cos{(\psi\!-\!\gamma)}\!
+\!\epsilon^2 \!\sin{2(\psi\!-\!\gamma)} \!\right\rangle
\!\left\langle \vphantom{L^2} \rho^{-1}\!
 \right\rangle \,,
$$
$$
D^*=\frac{1}{2} \left\langle \vphantom{L^2}
\! \epsilon^2\sin^2{(\psi\!-\!\gamma)}\!    \right\rangle
\left\langle \vphantom{L^2} \rho^{-2}
 \right\rangle  \,.
\eqno(29)
$$
Averages over $\psi$ reduce to constants due to stationarity
of $P(\psi)$.
The moments $\left\langle \vphantom{L^2} \rho^{m}  \right\rangle$
of the log-normal distribution (10) have exponential
behavior
$$
\left\langle \vphantom{L^2} \rho^{m}  \right\rangle \sim
\exp{(\kappa_m L)}
\eqno(30)
$$
with parameters
$$
\kappa_m= \left \{ \begin{array}{cc}
v m+Dm^2\,,\quad m>-v/2D \qquad ({\rm 31a})\\
-v^2/4D\,,\quad m<-v/2D \qquad  ({\rm 31b})
\end{array} \right. .
$$
In calculation of
$\left\langle \vphantom{L^2} \rho^{m}  \right\rangle$
one should take into account, that the log-normal distribution
(10) is valid not for arbitrary $\rho$, but only for $\rho\agt 1$;
in the first case, the result (31a)  would be valid without
restrictions (Fig.3).\,\footnote{\,The integrated function
$\rho^m P(\rho)$ after the change of variables $x=\ln \rho$
accepts the Gaussian form, valid only for $x\agt 1$. In the
case $m>-v/2D$, the Gaussian function is strongly localized
near its maximum situated at large positive $x$, so
restriction $x\agt 1$ is of no importance. In the case
$m<-v/2D$ the maximum of the Gaussian function goes to
large negative $x$, and the integral is determined by its tail
in the region $x\agt 1$; the proportionality
coefficient in Eq.30 depends on details of the
distribution $P(\rho)$ for $\rho\alt 1$, while the parameter
$\kappa_m$ is independent of them.}

\begin{figure}
\centerline{\includegraphics[width=2.3 in]{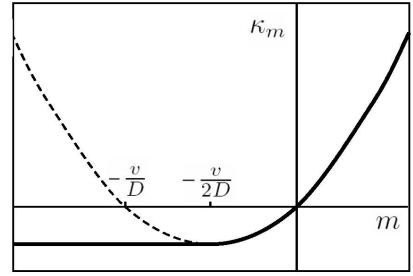}}
\caption{\small Parameter $\kappa_m$ in Eq.30 as a function of
$m$. The solid line is realized, if restriction $\rho\agt 1$
is accepted for the log-normal distribution (10), while
the dotted line corresponds to absence of such restriction
.  } \label{fig3}
\end{figure}

Since $\kappa_m$ are negative for negative $m$, it is
convenient to set
$$
\kappa_{-m} =- \tilde\kappa_m\,,\qquad m>0 \,,
\eqno(32)
$$
so equation for $P(\chi)$ accepts the  form
$$
\frac{\partial P}{\partial L} =
c_1 {\rm e}^{-\tilde\kappa_1 L} P'_\chi +
c_2 {\rm e}^{-\tilde\kappa_2 L} P''_{\chi \chi} \,,
\eqno(33)
$$
and can be solved iteratively for large  $L$,
$$
P_L(\chi) = P_\infty(\chi)
-\frac{c_1}{\tilde\kappa_1}{\rm e}^{-\tilde\kappa_1 L}
P_\infty'(\chi)
-\frac{c_2}{\tilde\kappa_2}{\rm e}^{-\tilde\kappa_2 L}
P_\infty''(\chi) \,,
\eqno(34)
$$
where $P_\infty(\chi)$ is the limiting distribution at
$L\to\infty$.

Let mark the points  $\tilde{\cal E}_1$ and $\tilde{\cal E}_2$
in Fig.1,b, corresponding to conditions $v=2D$ and $v=4D$. If the
log-normal distribution (10) was valid for arbitrary
$\rho$, then the striking phase transition
would occur at the point
$\tilde{\cal E}_1$, relating with sign reversal of $\tilde\kappa_2$
(the point $-v/D$ in Fig.3 coincide with $-2$ at
$\tilde{\cal E}=\tilde{\cal E}_1$, so
$\tilde\kappa_2>0$ for  $\tilde{\cal E}<\tilde{\cal E}_1$
and $\tilde\kappa_2<0$ for $\tilde{\cal E}>\tilde{\cal E}_1$).
Then the effective diffusion constant in Eq.33 would grow
with $L$ for $\tilde{\cal E}>\tilde{\cal E}_1$,
and would decrease for
$\tilde{\cal E}<\tilde{\cal E}_1$. For large $L$,
the distribution $P(\chi)$  would be homogeneous with high
accuracy for $\tilde{\cal E}>\tilde{\cal E}_1$, while the
non-trivial distribution $P_\infty(\chi)$ would be stabilized for
$\tilde{\cal E}<\tilde{\cal E}_1$.

\begin{figure}
\centerline{\includegraphics[width=2.1 in]{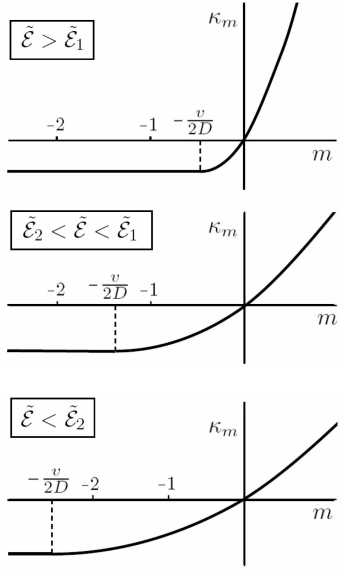}}
\caption{\small Mutual position of the points $-v/2D$,
$-1$ and $-2$ for $\tilde{\cal E}>\tilde{\cal E}_1$,
$\tilde{\cal E}_2<\tilde{\cal E}<\tilde{\cal E}_1$
and $\tilde{\cal E}<\tilde{\cal E}_2$.  }
\label{fig4}
\end{figure}

Due to restriction $\rho\agt 1$ such striking phase
transition is not realized\,\footnote{\,It is not excluded
that under special conditions the log-normal distribution
extends to the region $\rho\alt 1$, and this conclusion may
be revised.}, but the point $\tilde{\cal E}_1$
remains singular; analogous singularity arises at the
point $\tilde{\cal E}_2$. As should be clear from Fig.4,
the point $-v/2D$, corresponding to matching of the
parabola and constant, is situated on the right of the point
$-1$ for $\tilde{\cal E}>\tilde{\cal E}_1$, and
$$
\tilde\kappa_1=\tilde\kappa_2\quad \mbox{\rm for} \quad
\tilde{\cal E}>\tilde{\cal E}_1  \,.
\eqno(35)
$$
For the energy interval $\tilde{\cal E}_2<\tilde{\cal
E}<\tilde{\cal E}_1$, the point $-v/2D$ is located
between values $-2$ and $-1$, while for the interval
$\tilde{\cal E}<\tilde{\cal E}_2$ it appears on the left
of the point $-2$. One can see, that $\tilde\kappa_1$
has a jump of the second derivative at
$\tilde{\cal E}=\tilde{\cal E}_1$, while $\tilde\kappa_2$
has the analogous jump at $\tilde{\cal
E}=\tilde{\cal E}_2$.  These singularities can be easily
registered experimentally, using the treatment based on
Eq.34.  It is sufficient to find the limiting distribution
$P_\infty(\chi)$ and fit  $P_L(\chi)$  by dependence
$a P_\infty+bP'_\infty+cP''_\infty$: it is the linear fitting
procedure, which is easily realized by standard routines
\cite{20}. Condition (15) corresponds to a large
concentration of weak scatterers: in this case, coefficients in
Eq.27 changes slowly, which leads to formation of the Gaussian
distribution for $P(\chi)$ with variable
parameters\,\footnote{\,Above is valid in the case of
sufficiently strong localization of the distribution
$P(\chi)$; in the general case, it has a form of the sum of the
Gaussian functions, whose centers are separated by $2\pi$,
so $2\pi$-periodicity of solution is ensured.}. It is determined
 by the first two moments, which significantly simplifies a
treatment procedure.

\begin{center}
{\bf 6. Possibilities of measurements in single-mode waveguides }
\end{center}

\begin{center}
{\bf 6.1. Analogy with optics }
\end{center}

Localization of classical waves was discussed in a number of
papers  \cite{200}--\cite{206}, \cite{10,11}. It includes
consideration of  weak \cite{201} and  strong \cite{202,203}
localization, absorption near a photon mobility edge \cite{200},
near-field mapping of intensity of optical modes in
disordered waveguides \cite{205},
and many other  aspects (see the review article \cite{204}).
The transfer matrix approach to the problem was discussed in
\cite{10,11,206}. In application to optics the corresponding
analysis reduces to a set of simple relations.

Propagation of electromagnetic waves in homogeneous
dielectric media is described by the wave equation
$$
c^2 \Delta \Psi-n^2 \frac{\partial^2 \Psi}{\partial t^2}=0\,,
\eqno(36)
$$
where $\Psi$ is any component of the electric or magnetic field.
If a medium is spatially inhomogeneous, the refractive
index $n$ fluctuates along the coordinate $x$,

$$
n^2(x)=n_0^2+\delta
n^2(x)\,, \eqno(37)
$$
and for the monochromatic wave $\Psi\sim e^{i\omega t}$,
the wave equation can be written in the form
$$
\tilde c^2 \Delta \Psi+\left[\omega^2+\omega^2
\frac{\delta n^2(x)}{n_0^2}
 \right] \Psi=0 \,,\qquad \tilde c=c/n_0\,.
 \eqno(38)
 $$
The latter exhibits
the same structure, as the Schr${\rm\ddot o}$dinger equation
for an electron with energy ${\cal E}$ and mass $m$ in the random
potential $V(x)$.  One can easily establish
the correspondence
$$
{\cal E}\,\Longleftrightarrow\, \omega^2 \,,\quad
\frac{1}{2m}\,\Longleftrightarrow\, \tilde c^2 \,,\quad
V(x)\,\Longleftrightarrow \,-
\omega^2 \frac{\delta n^2(x)}{n_0^2} \,.
 \eqno(39)
 $$
A certain difference from the condensed matter physics is
related to the $\omega$ dependence of the effective potential
$V(x)$, which of little importance, if one is restricted by a
small frequency interval of the continuous spectrum.

\begin{figure}
\centerline{\includegraphics[width=3.2 in]{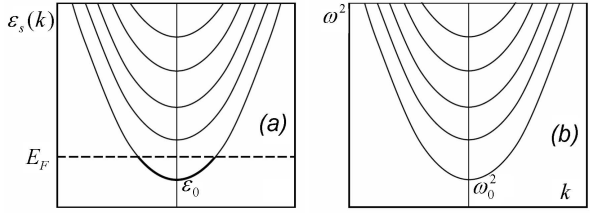}} \caption{
\small The spectrum of electrons in the metallic wire (a),
and the spectrum of waves in a metallic waveguide (b).  }
\label{fig5}
\end{figure}

The spectrum of waves propagating in a metallic waveguide is
analogous to the spectrum of electrons in a metallic
wire. In the latter case, the transverse motion is quantized,
leading to a set of the discrete levels $\epsilon_s$. If
the longitudinal motion is taken into account, these levels
transform to one-dimensional bands with the dispersion
law (Fig.5,a)
$$
\epsilon_s(k)=\epsilon_s + k^2/2m\,.
 \eqno(40)
$$
To obtain a strictly 1D system, one should have a
sufficiently small Fermi level so that only the lowest band
is occupied.
In the presence of impurities, the lower boundary $\epsilon_0$
of the spectrum is smeared out due to the appearance of fluctuation
states for  ${\cal E}<\epsilon_0$. The dependencies shown in Fig.1
correspond to the energy ${\cal E}$ counted from $\epsilon_0$.

Analogously, quantization of the transverse motion in
a metallic waveguide leads to a set of discrete frequencies
$\omega_s=\tilde c  \kappa_s$, where $-\kappa_s^2$  are
eigenvalues of the 2D Laplace operator with the appropriate
boundary conditions \cite{21}.
The zero eigenvalue is possible only in the case, when
the waveguide cross-section is multiply connected
(e.g. as in a coaxial cable).  For a singly connected
cross-section, the
minimum eigenvalue  $\omega_0$ is finite \cite{21}. If
the longitudinal motion is taken into account, the  following
branches of the spectrum are obtained (Fig.5,b)
$$
\omega^2_s(k)=\omega_s^2 + \tilde c^2 k^2\,.
 \eqno(41)
$$
To realize a single-mode regime, one should operate
near the lower boundary $\omega_0$ of the spectrum. In the
presence of disorder, the spectrum boundary $\omega_0$
is smeared out due to the occurrence of the fluctuation states.
Overall, the effects appearing in the electron system under the
change of the Fermi level can be observed in a single-mode
waveguide under the change of frequency $\omega$ in the vicinity
of $\omega_0$.

The spectrum in Fig.5,b corresponds to a metallic waveguide, which
is simply a hollow metal tube, which can be also filled by
non-absorptive dielectric. The latter case (a metal-coated
dielectric waveguide) is of the main interest for our purposes
due to possibility of addition of impurities providing
sufficiently strong elastic scattering. The coating thickness
should be of the order of the skin depth
in order to allow for partial field penetration (see Sec.6.4).
The transverse motion in the metallic waveguide is restricted by
the potential well with infinite walls, so multiplication by
$\omega^2$  (see (39)) has no effect, and parameters  $\kappa_s$
are constants, depending only on the form of the waveguide
cross-section; correspondingly, the spectrum in Fig.5,b is
strictly parabolic.

\begin{figure}
\centerline{\includegraphics[width=2.4 in]{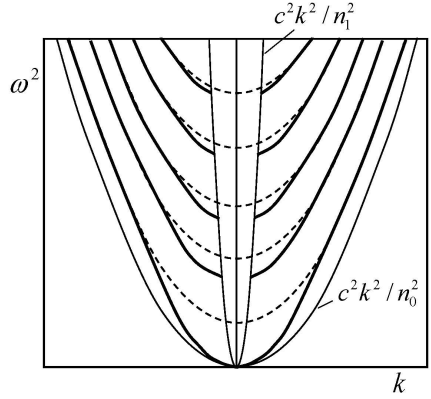}} \caption{
\small A spectrum of waves in a dielectric waveguide, with
the refractive index $n_0$ inside the waveguide and $n_1$ in
environment. For large $\omega$ the spectrum is the same as
in metallic waveguide (walls of the potential well are almost
infinite); if $\omega$ is diminished, then deviations arise
from the parabolic dependencies shown by dotted lines.
The lower restrictions for the allowed values of the longitudinal
momenta $k$ arise due to violation of conditions for the total
internal reflection. Disappearance of the boundary frequency
$\omega_0$ is related with the fact, that $\kappa_0^2$ is
restricted by the depth of the potential well,
proportional to $\omega^2$.  }
\label{fig6} \end{figure}

In the absence of metal coating (a pure dielectric waveguide),
the transverse motion is restricted by the potential well with
finite walls, and
the frequency dependence of the effective potential $V(x)$ (see
Eq.39) becomes essential.
Parameters  $\kappa_s$ cease to be constant and become $\omega$-dependent,
resulting in deviations from the parabolic dependencies in Fig.5,b.
In particular, the quantity $\kappa_0^2$ is restricted by the depth
of the potential, proportional to $\omega^2$, which leads to
disappearance of the boundary frequency $\omega_0$ (Fig.6).
In addition, the lower restrictions for the allowed
values of the longitudinal  momenta $k$ arise, related
with violation of conditions for the total internal reflection.
In the usual Schr${\rm\ddot o}$dinger equation, the bound states
in the potential well $V(x)$ correspond to the energy interval
$V_{min}<{\cal E}<V_\infty$, where $V_{min}$ is the minimal value of the
potential $V(x)$, and $V_\infty$ is its limiting (constant) value at infinity.
The corresponding condition in the dielectric waveguide has a form
$n_1^2 \omega^2 < c^2 k^2 < n_0^2 \omega^2$, where $n_0$ and $n_1$ are
refractive indices inside the waveguide and in its environment:
correspondingly, the spectrum of waves in the waveguide is
restricted by two parabolas (Fig.6).

One can see that a pure dielectric waveguide does not provide a
complete analogy with the electron disordered systems: there
is nothing that correspond to a
forbidden band, and certain differences occur near the band edge.
However, the allowed band is achievable for
investigation\,\footnote{\,Experimentally, the use of a pure
dielectric waveguide has certain advantages, relating with
absence of the Ohmic losses in metal coating.}: in
particular, the phase transition in $P(\psi)$ is situated in the
allowed band and may survive in a dielectric waveguide (though it
cannot be stated on a formal level). Its existence looks probable
for a sufficiently strong disorder, when the transition is
expected in the region where the actual spectrum is close to a
parabolic one.


\begin{center}
{\bf 6.2. Detection of phase transition in the $\psi$ distribution }
\end{center}

Let a wave of the unit amplitude be incident
from the left side of a single-mode waveguide,
and comes through it with the amplitude $t$, being
reflected with the amplitude $r$. If there are point scatterers
in the waveguide, then a partial reflection occurs at any of
them. Thus, at an arbitrary point $x$ of the waveguide one
finds a superposition of two waves, propagating in opposite
directions. The electric field $E(x,t)$
is determined by the  real part of this superposition, i.e.
$$
E(x,t)= {\rm Re} \left[Ae^{ikx+i\omega t}+Be^{-ikx+i\omega t}
\vphantom{L^2_2}  \right] \,.
\eqno(42)
$$
With the transfer matrix $T$ being defined by Eqs.(1,2),
the amplitudes of the transmitted and reflected waves
are determined by the expression
$$
\left ( \begin{array}{cc} A \\ B \end{array} \right)\,=
T\,\left ( \begin{array}{cc} t \\ 0 \end{array}\right)=
\left ( \begin{array}{cc} |t|\sqrt{\rho\!+\!1}
\, e^{i\varphi-i\varphi_0} \\
|t|\sqrt{\rho} \, e^{-i\theta-i\varphi_0}
\end{array}\right) \,,
\eqno(43)
$$
where $\rho$, $\varphi$, $\theta$ are $x$-dependent and
$t=|t|e^{-i\varphi_0}$ is accepted.
If the amplitude $|t|$ is sufficiently small, then
the quantity $\rho$ is large in the whole waveguide,
except the vicinity of its right end.
Then $|A|\approx |B|$, and Eq.42 gives in this approximation
$$
E(x,t)= \!
{\rm Re}\! \left[|A|\,e^{ikx+i\omega t+i\varphi-i\varphi_0}\!
+\! |B|\,e^{-ikx+i\omega t-i\theta-i\varphi_0}
\vphantom{L^2_2}\right]
$$
$$   \approx
2|A|\cos{\left(kx+\chi/2\right)}
\cos{\left(\omega t-\psi/2-\varphi_0\right)} \,,
\eqno(44)
$$
so the phase $\chi$ controls the coordinate dependence, while
$\psi$ controls the time dependence. The phases $\psi$ and
$\chi$ remain constant between scatterers, and change
abruptly when passing through a scatterer.
If the concentration of impurities is large, then $\psi$ and
$\chi$ change with $x$ practically continuously, having
random variations on the scale of the scattering length.

Since the field $E(x,t)$ can
be measured in principle,
both phases $\chi$ and $\psi$ are theoretically observable.
This is the fundamental difference from the condensed matter
physics, where a superposition of waves refers to a wave
function, and should be squared in modulus to obtain the
observable quantities: in this case the phase $\psi$
is unobservable in principle. This phase would
become unobservable in optics, if only the average
intensity could be measured
(it means that equation (44) is squared and averaged over time).
It is easy to verify, that this conclusion remains valid also
for $|A|\ne |B|$.

\begin{figure}
\centerline{\includegraphics[width=3.1 in]{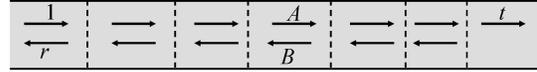}}
\caption{\small Propagation of waves in the
single-mode waveguide with point scatterers.  }
\label{fig7}
\end{figure}

Nevertheless, the occurrence
of the imaginary part of $\psi$
can be registered even in this case. If we suppose that
$$
\varphi=\varphi'+i\varphi''\,,\qquad
\theta=\theta'+i\theta''\,,
\eqno(45)
$$
then the amplitudes in the linear combination (42)
accept the form
$$
|A|= |t|\sqrt{\rho\!+\!1}
\, e^{-\varphi''} \,,\qquad
|B|=|t|\sqrt{\rho} \, e^{\theta''} \,.
\eqno(46)
$$
The flux conservation requires\,\footnote{\,A scattering is
considered as pure elastic. Inevitable Ohmic losses  in the
metal coating (see Sec.6.1) are suggested to be
essentially weaker in comparison with localization
effects. Sufficiently strong elastic scattering can be
provided in principle: e.g. in the case of the optical fibre
the impurity scattering is dominated for not very pure fibres
\cite{22}. }
that the condition $|A|^2=|B|^2+|t|^2$ be fulfilled at the
arbitrary point of the waveguide, which leads to relation
$$
(\rho\!+\!1)\, e^{-\varphi''} =
\rho \, e^{\theta''} +1
\eqno(47)
$$
giving $\theta''=-\varphi''$ for large $\rho$.
The imaginary part is absent in the phase $\chi$, but is
admissible for the phase $\psi$; in the latter case
$\psi''=2\theta''=-2\varphi''$, and in particular
$$
|A|= |t|\sqrt{\rho\!+\!1}
\, e^{\psi''/2} \,.
\eqno(48)
$$

The critical behavior of the imaginary part of $\psi$ can
be established from the general considerations. Let we
have the equation $F(x)=0$, where the function  $F(x)$
depends regularly on the external parameter $\epsilon$.
If at the point $\epsilon=0$ two real roots become
complex-valued, then the multiple root $x=p$ takes place
for $\epsilon=0$, and in its vicinity one has the equation
(the first derivative over $\epsilon$ is supposed to be finite)
$$
(x-p)^2-a\epsilon=0\,,
\eqno(49)
$$
which gives roots $p\pm \sqrt{a\epsilon}$ for $a\epsilon>0$
and roots $p\pm i \sqrt{|a\epsilon|}$ for $a\epsilon<0$.
Thereby, the appearance of the imaginary part is related
with a square root singularity.
According to Sec.3, the imaginary part of $\psi$ arises
in the result of a choice of parameters $s$ and $w_0$, providing
the correct values of $v$ and $D$ in the log-normal distribution
(10). Thereby, the parameters $s$ and $w_0$ are determined by
solution of certain equations, whose numerical analysis shows
\cite{17}, that appearance of the imaginary part of $w_0$
is related with confluence of two real roots and their
subsequent shift to the complex plane\,\footnote{\,The
second real root corresponds to the unphysical branch and
was not discussed in Ref.17. }. Hence, the above considerations
are immediately applicable to this situation:  if the
imaginary part of $\psi$ appears for $\omega<\omega_c$, then it
has a behavior\,\footnote{\,Usually in the phase transitions
theory, the square-root behavior of the order parameter
corresponds to the mean field theory, while the influence
of fluctuations leads to formation of the non-trivial
critical exponent $\beta$, which is less than 1/2. At
the present time, we do not see any indications for realization
of such scenario.}
$$
\psi''\sim\sqrt{\omega_c-\omega}\,\Theta(\omega_c-\omega)\,.
\eqno(50)
$$
According to  \cite{17},  the distribution $P(\rho)$ is not
singular at the point $\omega_c$ (Fig.1,b). It  refers to
a value of $\rho$ at the arbitrary point of the waveguide, and
in particular to its value at the whole length $L$,
which is related with $t$ as $|t|=(1+\rho)^{-1/2}$. Therefore,
the singularity in the amplitude (48) is completely
determined by the quantity $\psi''$ and has a square root
character. The square roots singularities at the point ${\cal E}_0$
are visually distinguishable in Figs.8,11 of the paper [17],
though obtained by numerical analysis.

\begin{figure}
\centerline{\includegraphics[width=3.4 in]{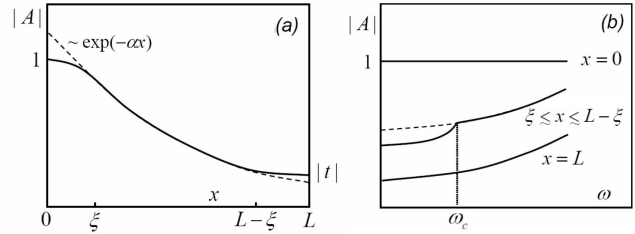}}
\caption{\small  (a) Dependence of the amplitude $|A|$
of the transmitted wave on the coordinate $x$ inside the
waveguide.  (b) The amplitude $|A|$ versus a frequency $\omega$
in the vicinity of the phase transition.  }
\label{fig8}
\end{figure}

The general picture looks as follows (Fig.8). In whole, the
modulus of $A$ changes in the waveguide according to the
exponential law, $|A|\sim e^{-\alpha x}$, but deviations from it
arise on the scale $\xi$ near the ends  due to influence of
boundary conditions (Fig.8,a): in particular, $|A|=1$ for $x=0$
and $|A|=|t|$ for $x=L$. The latter quantity is related with
$\rho$ and is a regular function of $\omega$.  However, in the
deep of the waveguide the amplitude $|A|$ has a square root
singularity (Fig.8,b), which can be registered already in
the measurements of the average intensity.
Such singularity can be observed at the specific point of
the system for a specific realization of the potential, since the
transition from the true transfer matrix to the pseudo one occurs
at the energy corresponding to the renormalized band edge shifted
due to a random potential\,\footnote{\,In this case, the square
root singularity can be obtained trivially from the behavior of
the true and pseudo transfer matrices for a point scatterer when
a shifted edge of the  band is approahed (see Ref.15).  }.
This shift changes from a point to point (see Footnote 2), but
for the distribution $P(\psi)$ in whole corresponds to a
strictly defined energy; the latter leads to square root
singularities for the moments of this distribution (see the
end of Sec.6.3).

According to \cite{17}, the critical point ${\cal E}_0$
is situated in the allowed band at the distance of order
$W^{4/3}$ from the band edge (Fig.1,b). Correspondingly,
in optics the critical point $\omega_c$
is greater than the boundary frequency $\omega_0$, while a distance between
them is determined by the degree of disorder.

\begin{center}
{\bf 6.3. Observability of phases $\psi$ and $\chi$
} \end{center}

Measurements of the time dependence
at optical frequencies are usually impossible. However,
observability of the phase $\psi$ can be
provided with heterodyne technique, in which
the measured electric field $E(x,t)$ is mixed with
the additional field $E_s(x,t)$, whose
frequency is shifted by a small quantity $\Omega$:
$$
E+E_s= {\rm Re} \left\{|E|e^{i\omega t+i\varphi_E}+
|E_s| e^{i(\omega+\Omega) t+i\varphi_s} \right\} \,.
\eqno(51)
$$
Considering the intensity averaged
over fast time oscillations, one has
$$
2\overline{(E+E_s)^2}=|E|^2+|E_s|^2+2|E||E_s|\cos{\left(\Omega
t+\varphi_s-\varphi_E\right)} \,,
\eqno(52)
$$
so the phase $\psi$ appears in combination with the
slow time dependendence, which can be measured by usual methods.
Substituting $E(x,t)$, corresponding to expression (44),
one obtains
$$
2\overline{(E+E_s)^2}=\left\{4|A|^2\cos^2{\left(kx+\chi/2\right)}
+|E_s|^2 \right\}+
$$
$$
+2|A|\cos{\left(kx+\chi/2\right)}\cdot
2|E_s|\cos{\left(\Omega t+\psi/2+\varphi_0+\varphi_s\right)}\,,
\eqno(53)
$$
so both phases $\chi$ and $\psi$ are observable, and
can be extracted from the experiment by the following treatment.

The stationary first term and the oscillatory second term in
Eq.53 can be separated by the Fourier analysis in the time domain.
The constant term $|E_s|^2$ can then be easily extracted,
since the smallest value of the first term in
the braces is zero. Since the cosine changes regularly and reverses
sign at any zero, the square root from the first
term in the braces can be extracted to inessential common sign.
As a result, two combinations would separately become known
$$
|A|\cos{\left(kx+\chi/2\right)} \,\,\, \mbox{\rm and} \,\,\,
|E_s|\cos{\left(\Omega t+\psi/2+\varphi_0+\varphi_s\right)} \,.
\eqno(54)
$$
The factor $|E_s|$ in the second combination is determined by
the amplitude of its temporal oscillations\,\footnote{\,Another
way to reach the same result is to make measurements
for several values of $|E_s|$ and fit the right-hand side of
Eq.53 by the dependence $\alpha +\beta |E_s| +\gamma
|E_s|^2$.}, while its $x$ dependence can be attributed to the
spatial dependence of the phase  $\psi$.

The treatment of the first combination (54) is complicated
by the fact that the amplitude $|A(x)|$ does not follow strictly
the exponential dependence $\exp(-\alpha x)$, but exhibits
significant fluctuations around it according to the
log-normal distribution (10). The appropriate treatment
looks as follows:

1. Find a value of $k$ by evaluating the average spatial
period of oscillations.

2. Find values of $\chi$ at the sequence of discrete
points, which are maxima, minima and zeroes of the oscillating
dependence, by assessing deviations of their position
from those of the purely cosine function. If the value
of $k$ is estimated correctly, then the obtained $\chi$ values
would fluctuate around a constant level
and not exhibit a systematic growth. As a result, one
can gather statistics
for the analysis of the $\chi$ distribution.

3. Find values of $|A(x)|$ at the points of maxima and
minima. These values would provide the data array
for verifying the log-normal distribution and  revelation
of systematic deviations from the exponential dependence
near the waveguide ends.

\vspace{2mm}

Observability of the phase $\psi$ provides additional
possibilities for registration of the phase transition.
If one introduce the variable $w$ defined in Eq.(11),
then the moments of the distribution $P(w)$ (e. g.
$\langle w \rangle$) will have the singularities
$\sqrt{\omega-\omega_c}$ in the region $\omega>\omega_c$. The
phase $\chi$ does not affect the evolution of $P(\rho)$ and
was not studied in the papers \cite{16,17}. However,
the possibility of its observation in optics makes such studies
to be actual.

\begin{center}
{\bf 6.4. The general measurement scheme }
\end{center}

\begin{figure}
\centerline{\includegraphics[width=2.1 in]{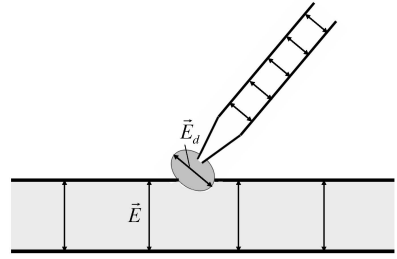}} \caption{
\small Measurement of the electrical field in a waveguide, using
the scanning near-field microscope in the detecting regime.
} \label{fig9}
\end{figure}

The electrical field in a waveguide can be measured
using methods of the scanning near-field optical microscopy
\cite{23,24,25}. There are two variants of the near-field
microscope, detecting and scattering, which determinate
two possible schemes of measurement. Comparison of these
schemes leads to a combined variant, where the problem of
detection reduces to the atomic force \cite{209,210} or tunneling
\cite{211} microscopy.

\vspace*{3mm}

{\it Detecting regime.} In this case, the probe of the
near-field microscope (a fragment of metal-coated
optical fibre)
is, in fact, a waveguide with the pointed tip and the
hole of sub-wavelength size $d$ (Fig.9). The near field, created
by the probe, can be imagined as a 'cloud' of finite volume
$V_d\sim d^3$ (see  Fig.4 in Ref.\cite{25}), with the
electrical field $\vec E_d$ approximately parallel to the field
inside the probe. Let the tip of the probe is approaching
at some angle to a surface of the given waveguide, so that
a certain volume $V$ of the 'cloud' penetrates inside the
waveguide  (Fig.9). If $\vec E$ is the measured field in
the waveguide, then a change of the energy  due to
penetration of the 'cloud' is determined by expression
$$
\left[ (\vec E+\vec E_d)^2 -\vec E^2 -\vec E_d^2\right]\,
V =2\vec E\cdot\vec E_d\, V \,.
\eqno(55)
$$
For small displacements $x$ of the probe, the change of the
volume is proportional to displacement, $\delta V=Sx$, where
$S$ is the square of intersection of the 'cloud' with
a surface of the waveguide. Then a force applied to the probe is
given by Eq.55 with replacement of $V$ by $S$. It can be
transformed to displacement of the probe, or
to the change of the
voltage retaining the probe in the fixed state. In fact, the
field $\vec E_d$ is space-dependent, and one should write
instead of (55)
$$
\int 2\vec
E\cdot\vec E_d(\vec r)\, d^3 r
\eqno(56)
$$
with integration over the waveguide volume, which reduces to
(55) after the rough estimation of the integral.

Assuming $E\sim E_d$, and introducing the atomic units of
the field strength and a force
$$
E_0=\frac{e}{a^2}\sim 10^9 \,\mbox{\it volt/cm}\,,\qquad
F_0=\frac{e^2}{a^2}\sim 10^{-2} \,\mbox{\it dyne}\,,
\eqno(57)
$$
we have the estimate of the force applied to the probe
$$
F\sim F_0 \left(\frac{E}{E_0}\right)^2
\left(\frac{d}{a}\right)^2\,.
\eqno(58)
$$
Since the size of the hole $d$ is restricted by the condition
$d\alt \lambda\sim 10^4 a$, we can set
$$
F\sim 10^{6} \left(\frac{E}{E_0}\right)^2
\,\mbox{\it dyne} \, .
\eqno(59)
$$
The maximal value of the field is restricted by the
field of the dielectric breakdown $\sim 10^7 volt/cm$.
Accepting sensitivity of measurement
on the level $F\sim 10^{-8} \mbox{\it dyne}$,
typical for the tunneling microscopy \cite{211},  we
have the wide interval of fields
$$
10^{-7} E_0 \alt E \alt 10^{-2} E_0\,,
\eqno(60)
$$
where the described scheme is realistic.

\begin{figure}
\centerline{\includegraphics[width=3.3 in]{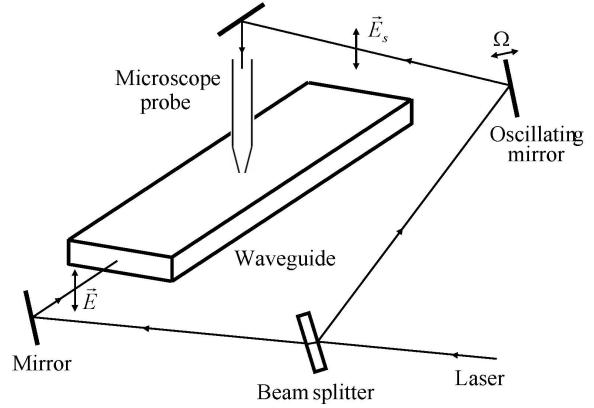}}
\caption{\small General scheme of measurement in the detecting
regime of the scanning near-field optical microscope.
} \label{fig10}
\end{figure}

If in the capacity of $\vec E_d$
we use the field $\vec E_s$ with a shifted
frequency (see Eq.51), then the force applied to the probe is
determined by the quantity
$$
F\sim S |A| |E_s|\cos{\left(kx+\chi/2\right)}
\cos{\left(\Omega t+\psi/2+\varphi_0+\varphi_s\right)}\,,
\eqno(61)
$$
whose treatment is even simpler then that of the
expression (53). In previous arguments, we did not take into
account existence of the semi-transparent metal coating (Sec.6.1)
and a difference from unity of the dielectric permeability inside
the waveguide. These factors leads to additive contribution
of order $E_s^2$ in the right-hand side of Eq.61, which is
independent of the measured field and easily separated in the
course of treatment.

The general measurement scheme looks as follows (Fig.10).
A laser beam is split into two parts, one of which is
directed into the waveguide. The second part of the
beam is incident to an oscillating mirror, acquiring a
small frequency shift $\Omega$ due to the Doppler effect.
Since the mirror velocity is variable, it leads to a variable shift
$\Omega$.  This problem can  be solved by registration of the
time dependence at the discrete points, equally spaced
by the period of mirror oscillations. Another possibility
consists in realization of the saw-toothed regime of
oscillations instead of the harmonic one.
Leaving the mirror, the beam is directed to the
microscope probe. Near a tip of the latter the field
$\vec E+\vec E_s$ is created, and measurement of the force
(61) allows to determine the coordinate dependence of
$\vec E$ in the course of scanning of the waveguide surface.

\vspace*{3mm}

{\it Scattering regime.} In this case, an optical microscope
probe is used not for the immediate field detection,
but only as a source of scattering\footnote{\,It can
be replaced by a needle of a scanning tunneling microscope,
which in the presence of metal coating
(see  Sec.6.1) allows one to use all
advantages of the scanning tunneling electron microscopy \cite{211}.  }
with subsequent use of a remote detector.
A wave propagating in the waveguide penetrates beyond its
boundaries due to the tunneling effect
and can be scattered by a probe tip
located close to a waveguide surface.  For sub-wavelength-sized probe tips,
the scattering occurs in the Rayleigh
regime, with the field of the scattered wave being
proportional to the local electric field $E(x,t)$ in the
waveguide\,\footnote{\,In the Rayleigh
scattering, the electromagnetic field of the scattered wave is
determined (in the main approximation) by the electric field of
the incident wave and does not depend on the wave vector of the
latter \cite{21}.  As a result, two waves entering the
superposition (42) are scattered equally, and the total field of
the scattered wave appears to be proportional to the electric
field in the waveguide.}
at the point of scattering $x$.

\begin{figure}
\centerline{\includegraphics[width=3.1 in]{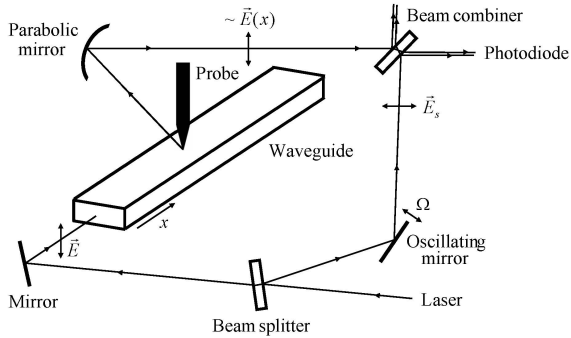}}
\caption{\small General scheme of measurement
in the scattering
regime of the scanning near-field optical microscope.
} \label{fig11}
\end{figure}

The general scheme of field measurement looks as follows (Fig.11).
A laser beam is split into two parts, one of which is
directed into a waveguide and eventually scattered by a
microscope probe tip.
The scattered light is collected by a parabolic mirror and
directed to a beam combiner. The second part of the laser beam is
reflected by an oscillating mirror, acquiring
a small frequency shift $\Omega$ due to the Doppler effect.
After the mirror, the beam is directed to the beam combiner,
where it is mixed with the first beam and
follows to a photodiode for measurement of intensity.
The described scheme was realized in studies of the paper
\cite{26}, where additional experimental details can be found.
\vspace*{3mm}

{\it Combined scheme} differs from Fig.10 only by the fact
that the second beam, leaving the oscillating mirror, is directed
to the waveguide and comes through it in the transverse direction
near its surface (Fig.12). Since the field $\vec E$ penetrates
beyond the waveguide due to the tunneling effect, the composed
field $\vec E+\vec E_s$ is present above its surface. The energy
of this field is changed, when the probe tip is
approached, due to the dielectric polarization of the latter.
As a result, the force applied to the probe is proportional to
the intensity  of the field $\vec E+\vec E_s$, and the problem of
its measurement is reduced to the atomic force \cite{209,210} or tunneling
\cite{211} microscopy.

\begin{figure}[b]
\centerline{\includegraphics[width=3.1 in]{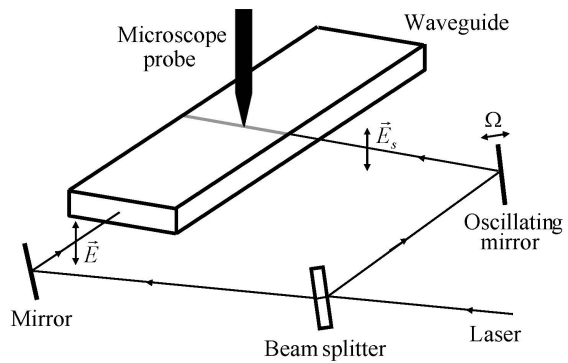}}
\caption{\small Measurement of the electric field in the
waveguide using the atomic force or tunneling microscopy.
} \label{fig12}
\end{figure}

\begin{center}
{\bf 7. Conclusion }
\end{center}

It has been shown that all results obtained for electrons
in 1D disordered systems are immediately applicable to the propagation
of electromagnetic waves in single-mode optical waveguides.
The modern optical methods enable  measurements of all parameters
$\rho$, $\psi$, $\chi$, entering the transfer
matrix. In particular, it becomes possible to observe the phase
transition for the distribution $P(\psi)$, which looks
unobservable in the framework of the condensed matter physics.
Since the phase $\chi$ becomes observable, one finds it actual to
derive the evolution equation for its distribution, which was not
studied in previous papers. At large $L$, the distribution of
$\chi$ has singularities, consisting in jumps of the second
derivative for exponentials, describing relaxation of $P_L(\chi)$
to the limiting distribution $P_\infty(\chi)$.

As was indicated above, one of the measurement schemes
described in Sec.6.4 was realized in the
paper \cite{26}. In contrast to studies \cite{207,208}, where
only the transmission matrix was measured, Ref.\cite{26}
presents the experimental approach  allowing to measure the
phase distribution inside the waveguide.  However, the
measurements of Ref.\cite{26} were not concerned with light
propagation in disordered systems, but only with characterization
of regular modes in homogeneous waveguides.

Essentially new measurements are necessary for testing the
validity of claims made in the present paper. The actual
experiments should be executed using a tunable laser allowing to
change the light frequency, and its tunability range should cover
the targeted phase transition. The latter demand to establish
the most promising waveguide configuration and dimensions.
A suitable approach should be developed for introducing a large
concentration of impurities into the waveguide.
Extensive analysis is necessary to find the parameter range, where
the localization effects will be dominating over the light
absorption inside the waveguide and radiative
losses through its boundaries. The latter problem is somewhat
facilated for a pure dielectric waveguide, but then the
analogy with electronic systems becomes incomplete (Sec.6.1).

One can hope that the obtained  results might stimulate
the corresponding experimental activities that would, in turn,
shed more light on intricate effects in both optical and
electron localization phenomena.

The author is indebted to S.I.Bozhevolnyi for numerous discussions
of the optical aspects of this paper.

\begin{center}
{\it Appendix.}{\bf \,\,
Evolution equation for  $P(\rho,\psi,\chi)$}
\end{center}

The method for derivation of the evolution equation presented
below is somewhat different from that in the papers
\cite{16,17}: it is more systematic  and ensures attainment
of a result when its character is unknown beforehand. The
more compact way of derivation \cite{16,17} can be found
only in the presence of a certain information on the structure
of the result.

As clear from relations (18), (19), (20), (22), the phase
$\psi$ enters evolution equations in the form of two
combinations $\psi-\gamma$ and $\psi-\beta$, so the shift
$\psi\to\psi+\psi_0$ allows to reduce the parameter
$\gamma$ to the value $-\pi/2$, corresponding to abrupt
interfaces between the system and the ideal leads \cite{17};
to simplify formulas, we restrict ourselves by this case.
Relations (12--14) give in the main order in $\delta$
$$
T^{(n+1)}_{11}= (1+i\alpha-i\epsilon_n) T^{(n)}_{11} +
(\delta_1-i\delta_2+i\epsilon_n) T^{(n)}_{12}
$$
$$
T^{(n+1)}_{12}= (\delta_1+i\delta_2-i\epsilon_n) T^{(n)}_{11}
+ (1-i\alpha+i\epsilon_n) T^{(n)}_{12}
\eqno(A.1)
$$
and analogous equations for $T^{(n)}_{21}$ and $T^{(n)}_{22}$,
obtained by complex conjugation; here  $\delta_1=\Delta
\cos\beta$, $\delta_2=\Delta \sin\beta$. Setting
$$
T^{(n)}_{11}=x_n+iy_n\,,
\qquad T^{(n)}_{12}=z_n+iw_n\,,
\eqno(A.2)
$$
we have
%
$$
x_{n+1}=x_n-(\alpha -\epsilon_n)y_n +\delta_1 z_n
-(\delta_2+\epsilon_n) w_n
$$
$$
y_{n+1}=(\alpha -\epsilon_n) x_n +y_n+(\delta_2 +\epsilon_n)z_n
+ \delta_1 w_n
$$
$$
z_{n+1} = \delta_1 x_n +(\delta_2 +\epsilon_n) y_n
+z_n + (\alpha-\epsilon_n) w_n
 \eqno(A.3)
$$
$$
w_{n+1}= -(\delta_2 +\epsilon_n) x_n +\delta_1 y_n
 -(\alpha- \epsilon_n)z_n +w_n
$$
which after rewriting in the matrix form gives the matrix
with the unit determinant. If the distribution
$P_n(x_{n},y_{n},z_{n},w_{n})$ is known, then the analogous
distribution at the $(n+1)$th step is composed according
to the rule
$$
P_{n+1}(\tilde x_{n+1},\tilde y_{n+1},\tilde z_{n+1},\tilde
w_{n+1}) = \int d\epsilon_n dx_{n}\,dy_{n}\,dz_{n}\,dw_{n}
\cdot
$$
$$   \cdot
P(\epsilon_n)\,P_n(x_{n},y_{n},z_{n},w_{n})
\delta\left(\tilde x_{n+1}-x_{n+1}\right)
 \cdot
\eqno(A.4)
$$
$$ \cdot
\delta\left(\tilde y_{n+1}-y_{n+1}\right)
\delta\left(\tilde z_{n+1}-z_{n+1}\right)
\delta\left(\tilde w_{n+1}-w_{n+1}\right)\,,
$$
where $x_{n+1}$, $y_{n+1}$, $z_{n+1}$, $w_{n+1}$ are
expressed in terms of  $x_{n}$, $y_{n}$, $z_{n}$,
$w_{n}$ according to $(A.3)$. Let inverse the relation
$(A.3)$ and come to integration over
$x_{n+1}$, $y_{n+1}$, $z_{n+1}$, $w_{n+1}$; since the
Jacobian is equal to unity and $\delta$-functions are
trivially removed, we come to equation
$$
P_{n+1}(x_{n+1},y_{n+1},z_{n+1},w_{n+1})=
\qquad \qquad\qquad
$$
$$ \qquad\qquad\qquad
=\int d\epsilon_n P(\epsilon_n)
P_n(x_{n},y_{n},z_{n},w_{n}) \,,
\eqno(A.5)
$$
where $x_{n}$, $y_{n}$, $z_{n}$, $w_{n}$ are expressed
through  $x_{n+1}$, $y_{n+1}$, $z_{n+1}$,
$w_{n+1}$ by the relation inverse to $(A.3)$. Expanding over
differences  $x_{n+1}-x_{n}$, $y_{n+1}-y_{n}$,
$\dots$ and retaining the terms of the first order in
 $\delta$ and second order in $\epsilon$, we have
$$
\frac{\partial P}{\partial n}= \alpha\left[
y\frac{\partial P}{\partial x} -x \frac{\partial P}{\partial y}
-w \frac{\partial P}{\partial z}
+z \frac{\partial P}{\partial w} \right] -
$$
$$
-\delta_1\left[ z\frac{\partial P}{\partial x}
+w \frac{\partial P}{\partial y}
+x \frac{\partial P}{\partial z}
+ y \frac{\partial P}{\partial w} \right] +
$$
$$
+\delta_2\left[ w\frac{\partial P}{\partial x}
-z \frac{\partial P}{\partial y}
-y \frac{\partial P}{\partial z}
+ x \frac{\partial P}{\partial w} \right] +
$$
$$
+ \frac{1}{2}\epsilon^2 (w\!-\!y)^2
\left[ \frac{\partial^2 P}{\partial x^2}
+2 \frac{\partial^2 P}{\partial x \partial z}
+\frac{\partial^2 P}{\partial z^2}  \right] +
\eqno(A.6)
$$
$$
+ \frac{1}{2}\epsilon^2 (x\!-\!z)^2
\left[ \frac{\partial^2 P}{\partial y^2}
+2 \frac{\partial^2 P}{\partial y \partial w}
+\frac{\partial^2 P}{\partial w^2}  \right] +
$$
$$
+ \epsilon^2 (x\!-\!z)(w\!-\!y)
\left[  \frac{\partial^2 P}{\partial x \partial y}
+\frac{\partial^2 P}{\partial x \partial w}
+\frac{\partial^2 P}{\partial z \partial y}
+\frac{\partial^2 P}{\partial z \partial w}  \right] \,.
$$
Introducing the polar coordinates
$$
x=r_1 \cos\varphi,\quad y=r_1\sin\varphi,
\quad z=r_2 \cos\theta, \quad w=r_2 \sin\theta\,,
\eqno(A.7)
$$
we obtain
$$
\frac{\partial P}{\partial n}=
\alpha\left[ -P'_\varphi+P'_\theta  \right]
-\Delta \cos(\theta-\varphi-\beta)
\left[ r_2 P'_{r_1}+r_1 P'_{r_2}  \right] +
$$
$$
+\Delta \sin(\theta\!-\!\varphi\!-\!\beta)
\left[ \frac{r_1}{r_2} P'_{\theta}-
\frac{r_2}{r_1} P'_{\varphi}   \right] +
$$
$$
+ \frac{1}{2}\epsilon^2 \left\{
\vphantom{\left[ \frac{1}{r^2_2} P''_{\theta \theta} \right]}
\sin^2(\theta\!-\!\varphi)
\left[ r^2_2 P''_{r_1 r_1}+2 r_1 r_2 P''_{r_1 r_2}
+ r^2_1 P''_{r_2 r_2} \right] \right. +
$$
$$ \left.
+ 2\sin(\theta\!-\!\varphi)
\left[ \vphantom{r^2_2 P''_{r_1 r_1}}
         r_1-r_2\cos(\theta\!-\!\varphi)  \right]
\left[ \frac{r_2}{r_1} P''_{r_1 \varphi} +P''_{r_2 \varphi}
-\frac{r_2}{r^2_1} P'_{ \varphi} \right] \right.
$$
$$ \left.
+ 2\sin(\theta\!-\!\varphi)
\left[ \vphantom{r^2_2 P''_{r_1 r_1}}
       r_1\cos(\theta\!-\!\varphi)-r_2  \right]
\left[ \frac{r_1}{r_2} P''_{r_2 \theta}+P''_{r_1 \theta} -
\frac{r_1}{r^2_2} P'_{ \theta} \right]  \right.
$$
$$
+\left[ \vphantom{r^2_2 P''_{r_1 r_1}}
      r_1-r_2\cos(\theta\!-\!\varphi)  \right]^2
\left[ \frac{1}{r^2_1} P''_{\varphi \varphi}
       +\frac{1}{r_1} P'_{ r_1} \right]+
$$
$$
+\left[\vphantom{r^2_2 P''_{r_1 r_1}}
      r_1\cos(\theta\!-\!\varphi)-r_2  \right]^2
\left[ \frac{1}{r^2_2} P''_{\theta \theta}
       +\frac{1}{r_2} P'_{ r_2} \right] +
\eqno(A.8)
$$
$$ \left.
+2\left[ \vphantom{r^2_2 P''_{r_1 r_1}}
      r_1-r_2\cos(\theta\!-\!\varphi)  \right]
\left[ \vphantom{r^2_2 P''_{r_1 r_1}}
      r_1\cos(\theta\!-\!\varphi)-r_2  \right]
\frac{1}{r_1 r_2} P''_{\varphi \theta}
\vphantom{\left[ \frac{1}{r^2_2} P''_{\theta \theta} \right]}
                                  \right\}  \,.
$$
Now come from the quantities $r_1$, $r_2$  to the new variables
$\rho$, $\xi$
$$
 r^2_1+r^2_2 =1+2\rho\,,\qquad
 r^2_1-r^2_2 =\xi\,.
\eqno(A.9)
$$
On can easily verify, that all terms with derivatives
over $\xi$ disappear; hence the quantity $\xi$ remains
constant in the course of evolution, and on the physical
grounds we can set $\xi=1$.  Then
$$
r_1=\sqrt{1+\rho}\,,\qquad r_2=\sqrt{\rho}
\eqno(A.10)
$$
in correspondence with the canonical representation (2).
The corresponding evolution equation accepts the form
$$
\frac{\partial P}{\partial n}=
\alpha\left[ -P'_\varphi+P'_\theta  \right]
-\Delta \cos(\theta\!-\!\varphi\!-\!\beta)
2 r_1 r_2 P'_{\rho} +
$$
$$
+\Delta \sin(\theta\!-\!\varphi\!-\!\beta)
\left[ \frac{r_1}{r_2} P'_{\theta}-
\frac{r_2}{r_1} P'_{\varphi}   \right] +
$$
$$
+ \frac{1}{2}\epsilon^2 \left\{
\vphantom{\left[ \frac{1}{r^2_2} P''_{\theta \theta} \right]}
4 r^2_1 r^2_2 \sin^2(\theta\!-\!\varphi) P''_{\rho \rho}+
\right.
$$
$$
+\left[ 2 r^2_1 +2 r^2_2-4 r_1 r_2 \cos(\theta\!-\!\varphi)  \right]
P'_{ \rho}+
$$
$$
+ 4r_2\sin(\theta\!-\!\varphi)
\left[ \vphantom{r^2_2 P''_{r_1 r_1}}
    r_1-r_2\cos(\theta\!-\!\varphi)
    \right] P''_{\rho \varphi} +
$$
$$
 + 4r_1\sin(\theta\!-\!\varphi)
 \left[ \vphantom{r^2_2 P''_{r_1 r_1}}
      r_1\cos(\theta\!-\!\varphi)-r_2  \right] P''_{\rho \theta}
-
$$
$$
-2\sin(\theta\!-\!\varphi)
\frac{r_2\left[ r_1-r_2 \cos(\theta\!-\!\varphi) \right]}{r_1^2}
P'_{ \varphi} -
\eqno(A.11)
$$
$$
-2\sin(\theta\!-\!\varphi)
\frac{r_1\left[ r_1\cos(\theta\!-\!\varphi)-r_2  \right]}{r_2^2}
P'_{ \theta}
+
$$
$$ \left.
+\!\left[ \frac{r_1 \!-\!r_2\cos(\theta\!-\!\varphi)}{r_1}
\right]^2 \!\! P''_{\varphi \varphi} + \! \left[
\frac{r_1\cos(\theta\!-\!\varphi)\!-\!r_2}{r_2} \right]^2 \!\!
P''_{\theta \theta} + \right.
$$
$$ \left.  +2\left[
\frac{r_1-r_2\cos(\theta\!-\!\varphi)}{r_1} \right]
\left[\frac{r_1\cos(\theta\!-\!\varphi)-r_2}{r_2} \right]\,
P''_{\varphi \theta}
\vphantom{\left[ \frac{1}{r^2_2} P''_{\theta \theta} \right]}
                                  \right\}  \,.
$$
In the course of the changes of variables $(A.7)$ and
$(A.9)$
we do not produce renormalization of probability; however,
in the result of two changes we have
$$
4 P(x,y,z,w)\,dx \,dy \,dz \,dw = P(\rho,\xi,\varphi,\theta)
\,d\rho \,d\xi \,d\varphi \,d\theta  \,,
\eqno(A.12)
$$
and the indicated renormalization reduces to an inessential
constant factor. Introducing combined phases (3), one has
$$
\frac{\partial P}{\partial n}=
2\alpha P'_\psi
-\Delta \cos(\psi\!-\!\beta) 2 r_1 r_2 P'_{\rho}+
$$
$$
+\Delta \sin(\psi\!-\!\beta)
\left[ \left( \frac{r_1}{r_2}+ \frac{r_2}{r_1} \right) P'_{\psi}
+\left( \frac{r_1}{r_2}- \frac{r_2}{r_1} \right) P'_{\chi}
\right] +
$$
$$ +
\frac{1}{2}\epsilon^2 \left\{
\vphantom{\left[ \frac{1}{r^2_2} P''_{\theta \theta} \right]}
4 r^2_1 r^2_2 \sin^2{\psi} P''_{\rho \rho}
+\left[ 2 r^2_1 +2 r^2_2-4 r_1 r_2 \cos{\psi}
\right] P'_{\rho}
\right. +
$$
$$
+ 4\sin{\psi} \left(
\vphantom{r^2_2 P''_{r_1 r_1}} r_1 r_2-r_2^2\cos{\psi}  \right)
 \left( - P''_{\rho \psi} +P''_{\rho \chi}     \right)
 $$
 $$
+ 4\sin{\psi} \left(
\vphantom{r^2_2 P''_{r_1 r_1}} r_1^2\cos{\psi}-r_1 r_2  \right)
 \left( P''_{\rho \psi} +P''_{\rho \chi}     \right)
$$
$$
- 2\sin{\psi} \left( \frac{r_2}{r_1}-
     \frac{r^2_2}{r^2_1} \cos{\psi}  \right)
 \left( - P'_{\psi} +P'_{ \chi}     \right)
\eqno(A.13)
 $$
 $$
- 2\sin{\psi} \left( \frac{r^2_1}{r^2_2} \cos{\psi}
     -\frac{r_1}{r_2} \right)
     \left( P'_{\psi} +P'_{ \chi}  \right)
$$
$$
+\left( \frac{r_1-r_2\cos{\psi}}{r_1} \right)^2
\left( P''_{\psi \psi}- 2P''_{\psi \chi}+P''_{\chi \chi}
\right)
$$
$$
+\left( \frac{r_1\cos{\psi}-r_2}{r_2} \right)^2
\left( P''_{\psi \psi}+2 P''_{\psi \chi}+P''_{\chi \chi}
\right)
$$
$$
\left.
+\left( \frac{r_1-r_2\cos{\psi}}{r_1} \right)
\left( \frac{r_1\cos{\psi}-r_2}{r_2} \right)
\left( -P''_{\psi \psi}+P''_{\chi \chi}
\right)     \right \}
$$
Substituting $(A.10)$ and transforming the right-hand side
to the sum of full derivatives, we come to the final evolution
equation, which has a structure of Eq.4:
$$
\frac{\partial P}{\partial n}=
\left\{\vphantom{\frac{1}{2}}
\!\!-2\Delta\!\cos{(\psi\!-\!\beta)} \sqrt{\!\rho(\!1\!+\!\rho\!)} P+
2\epsilon^2\! \sin\!^2{\psi} \,\rho(\!1\!+\!\rho\!)  P'_{\rho}+
\right.
$$
$$
+\epsilon^2\left[(1\!-\!2\sin^2{\psi})(1\!+\!2\rho)
   -2\cos{\psi}\sqrt{\rho(1\!+\!\rho)} \right] P +
$$
$$ \left.
+2\epsilon^2 \sin{\psi}\left[ \cos{\psi}(1\!+\!2\rho)
   -2\sqrt{\rho(1\!+\!\rho)} \right] P'_{\psi} \right\}'_\rho +
$$
$$
+\left\{\left[ \vphantom{R^2}
2\alpha\!+\!R\Delta\sin{(\psi\!-\!\beta)}\right] P
\vphantom{\frac{1}{2}} +\epsilon^2\sin{\psi}(R\!-\!2\cos{\psi}) P
 \right.+
$$
$$ \left.
+\frac{1}{2}\epsilon^2 (2\!-\!R\cos{\psi})^2\, P'_\psi
 \right\}'_\psi \,+
\eqno(A.14)
$$
$$
+\left\{\,\frac{ \Delta\sin{(\psi\!-\!\beta)}\!
 +\!\epsilon^2\sin{\psi}\, (1\! -\! R \cos{\psi}) }
{\sqrt{\rho(1\!+\!\rho)}}\, P  \right.
$$
$$ \left.
+\, \frac{ \epsilon^2 \cos{\psi}\, (R \cos{\psi}\! -\! 2) }
{\sqrt{\rho(1\!+\!\rho)}}\, P'_\psi
+ \frac{ \epsilon^2\cos^2{\psi} }
{2\rho(1\!+\!\rho)}\, P'_\chi \,
 \right\}'_\chi \,.
$$
Integration over $\chi$ gives the evolution equation
for $P(\rho,\psi)$, obtained in \cite{16,17}, while integration
over $\rho$ and $\psi$  leads to equation (27)  for $P(\chi)$.


\begin{thebibliography}{xx}

\bibitem{1} P. W. Anderson, D. J. Thouless, E. Abrahams,
D. S. Fisher, Phys. Rev. B {\bf 22}, 3519 (1980).


\bibitem{2} R. Landauer, IBM J. Res. Dev. {\bf 1}, 2 (1957);
Phil. Mag. {\bf 21}, 863 (1970).

\bibitem{3} V. I. Melnikov,
Sov. Phys. Sol. St.  {\bf 23}, 444 (1981).

\bibitem{4} A. A. Abrikosov, Sol. St. Comm. {\bf 37},
997 (1981).

\bibitem{5} N. Kumar, Phys. Rev. B {\bf 31}, 5513 (1985).

\bibitem{6} B. Shapiro, Phys. Rev. B {\bf 34}, 4394 (1986).

\bibitem{7} P. Mello, Phys. Rev. B {\bf 35}, 1082 (1987).

\bibitem{8} B. Shapiro, Phil. Mag. {\bf 56}, 1031 (1987).

\bibitem{9}  I. M. Lifshitz, S. A. Gredeskul, L. A. Pastur,
Introduction to the Theory of Disordered Systems, Nauka, Moscow,
1982.

\bibitem{10} C. W. J. Beenakker, Rev. Mod. Phys.
{\bf 69}, 731 (1997).

\bibitem{11} X. Chang, X. Ma, M. Yepez, A. Z. Genack, P. A.
Mello,  Phys. Rev. B {\bf 96}, 180203 (2017).

\bibitem{12} L. I. Deych, D. Zaslavsky, A. A. Lisyansky,
 Phys. Rev. Lett. {\bf 81}, 5390 (1998).

\bibitem{13} L. I. Deych, A. A. Lisyansky, B. L Altshuler,
Phys. Rev. Lett. {\bf 84}, 2678 (2000); Phys. Rev. B
{\bf 64}, 224202 (2001).

\bibitem{14} L. I. Deych, M. V. Erementchouk, A. A. Lisyansky,
 Phys. Rev. Lett. {\bf 90}, 126601 (2001).

\bibitem{15}   I.\,\,M.\,\,Suslov,
J. Exp. Theor. Phys. {\bf 129}, 877 (2019)
[Zh. Eksp. Teor. Fiz. {\bf 156}, 950 (2019)].

\bibitem{16} I. M. Suslov, Phil. Mag. Lett.  {\bf 102}, 255
(2022).

\bibitem{17}   I.\,\,M.\,\,Suslov,
J. Exp. Theor. Phys. {\bf 135}, 726 (2022)
[Zh. Eksp. Teor. Fiz. {\bf 162}, 750 (2022)].

\bibitem{300} S. I. Bozhevolnyi, I. M. Suslov,
Phys. Scr. {\bf 98}, 065024 (2023).


\bibitem{18}   I.\,\,M.\,\,Suslov, Adv. Theor. Comp. Phys. 6, 77
 (2023).

\bibitem{19}   N. F. Mott, E. A Davis, Electron Processes in
Non-Crystalline Materials, Oxford, Clarendon Press, 1979.

\bibitem{122} V. V. Brazhkin, I. M. Suslov,
J. Phys. -- Cond. Matt.  32 (35), 35LT02 (2020).

\bibitem{123}  I.\,\,M.\,\,Suslov,
J. Exp. Theor. Phys. {\bf 131}, 793 (2020)
[Zh. Eksp. Teor. Fiz. {\bf 158}, 911 (2020)].



\bibitem{200} S. John, Phys. Rev. Lett. 53, 2169 (1984).

\bibitem{201} P. Van Albada, A. Lagendijk, Phys. Rev. Lett. 55,
2692 (1985).

\bibitem{202}  P.  W.  Anderson,  Phil.  Mag.  B 52,  505
(1985).

\bibitem{203}  S.  John,  Phys.  Rev.  Lett.  58,  2486 (1987).

\bibitem{205} S. I.  Bozhevolnyi, V. S. Volkov, K. Leosson, Phys.
Rev.  Lett.  89, 186801  (2002).


\bibitem{204} D. S. Wiersma, Nature Photon. 7, 188 (2013).


\bibitem{206} Zh. Shi, M. Davy, A. Z. Genack,  Opt. Express 23,
 12293 (2015).






\bibitem{20} W. H. Press, B. P. Flannery, S. A. Teukolsky,
W. T. Wetterling, Numerical Recipes in Fortran, Cambridge
University Press, 1992.



\bibitem{21} L. D. Landau, E. M. Lifshits,
Electrodynamics of Continuous Media,
Oxford, Pergamon Press, 1984.


\bibitem{22}  Ch. K. Kao, Nobel Prize Lecture, 2009.

\bibitem{23} D. W. Pohl, W. Denk, M. Lanz, Appl. Phys. Lett.
{\bf 44}, 651 (1984).

\bibitem{24} D. W. Pohl, L. Novotny, J. Vac. Sci. Technol.
B  {\bf 12}, 1441 (1994).

\bibitem{25} A. L. Lereu, A. Passian, Ph. Dumas, Int. J.
Nanotechnol. {\bf 9}, 488 (2012).

\bibitem{211} G.Binning, H.Rohrer, Helv. Phys. Acta.
{\bf 55}, 726 (1982).


\bibitem{209} G.Binning, C. F. Quate, C. Gerber, Phys. Rev. Lett.
{\bf 56}, 930 (1986).

\bibitem{210} E. Meyer, Progress in Surface Science {\bf 41}, 3
(1992).



\bibitem{26} S. I. Bozhevolnyi, V. A. Zenin, R. Malreanu,
I. P. Radko, A. V. Lavrinenko,  Opt. Express 24, 4582 (2016).

\bibitem{207} I. M. Vellekoop and A. P. Mosk, Phys. Rev. Lett. 101, 120601 (2008).

\bibitem{208} S. M. Popoff, G. Lerosey, R. Carminati, M. Fink, A. C.
Boccara, and S. Gigan, Phys. Rev. Lett. 104, 100601 (2010).


\end{thebibliography}
\end{document}